\documentclass[pre,showpacs,preprint,superscriptaddress]{revtex4}
\usepackage{amssymb,graphicx,amsbsy}

\begin{document}

\title{Comment on `Six-state clock model on the square lattice: Fisher zero
approach with Wang-Landau sampling'}
\author{Seung Ki Baek}
\affiliation{Department of Physics, Ume{\aa} University, 901 87 Ume{\aa}, Sweden}
\author{Petter Minnhagen}
\affiliation{Department of Physics, Ume{\aa} University, 901 87 Ume{\aa}, Sweden}
\author{Beom Jun Kim}
\email[Corresponding author, E-mail: ]{beomjun@skku.edu}
\affiliation{BK21 Physics Research Division and Department of Physics, 
Sungkyunkwan University, Suwon 440-746, Korea}

\begin{abstract}
Hwang in [Phys. Rev. E {\bf{80}}, 042103 (2009)] suggested that the two transitions of the six-state clock model on the square lattice are
\emph{not} of the Kosterlitz-Thouless type. Here we show from simulations that
at the upper transition, the helicity modulus does make a discontinuous jump
to zero. This
gives strong evidence for a Kosterlitz-Thouless
transition.
\end{abstract}


\maketitle

Recently, Hwang~\cite{hwang} has examined the six-state clock model by the
Fisher-zero approach and suggested that its phase transitions are not of the
Kosterlitz-Thouless (KT) type, in contrary to earlier theoretical
analyses~\cite{jkkn,elit}. This raises questions of earlier
numerical works where scaling indices have been measured and found to be 
in agreement with the earlier theoretical KT predictions (see, for example,
Ref.~\cite{tomita}).
In spite of the theoretical and numerical supports for the KT scenario
in the six-state clock model, one might perhaps still argue that the
actual transitions can nevertheless be of a standard continuous type due to
the following reasons.
First, the theoretical predictions usually relate to the Villain
approximation which is not an exact representation of the clock model.
Second, the agreements with the KT scaling indices only provide a necessary
condition but maybe not a sufficient one for ruling out a standard
continuous transition.
Third, it has been claimed in Ref.~\cite{lapilli} that the six-state clock
model does not exhibit a discontinuous jump in the helicity modulus $\left<
\Upsilon \right>$, a key feature of the KT behavior. Thus, if this is
correct, one can definitely rule out a KT transition.  However, as shown
here, it is in fact not correct: in this
Comment, we present simulation results for the fourth-order helicity modulus
$\left< \Upsilon_4 \right>$ and from the numerical results we directly
verify the discontinuous character of the helicity modulus at the upper
phase transition of the six-state clock model.

Reference~\cite{petter} proposed a numerical method to identify the
KT transition in the two-dimensional $XY$ model based on a stability
argument. That is, an external twist of magnitude $\Delta$ across the
$XY$-spin system gives an additional contribution to the free energy $F$ so
that $F(\Delta) \ge F(\Delta=0)$. For the system to be stable under small
$\Delta$, where one can expand the free energy density $f$ ($\equiv F/N$ with
$N$ being the number of $XY$ spins) as $f(\Delta) = \left<
\Upsilon \right> \frac{\Delta^2}{2} + \left< \Upsilon_4 \right>
\frac{\Delta^4}{4!} + \cdots$, the helicity modulus $\left< \Upsilon
\right>$ must be non-negative. In the $XY$ model, this quantity is zero
above the critical temperature $T_{KT}$ and positive finite below $T_{KT}$
in the thermodynamic limit. The same stability consideration tells us that
the fourth-order helicity modulus $\left< \Upsilon_4 \right>$ also must be
nonnegative at any temperature $T$ where $\left< \Upsilon \right>$ vanishes.
Suppose that $\left< \Upsilon_4 \right>$ is finite and negative at the
transition. Then $\left< \Upsilon \right>$ cannot approach zero
continuously but must instead make a discontinuous jump to zero at
the transition. Such a discontinuous jump is precisely the characteristic
behavior of $\left< \Upsilon \right>$ at the critical temperature for a
KT transition. Unfortunately, it is notoriously difficult to verify the
discontinuous character of the helicity modulus directly from numerical
simulations because the precision of the simulations are restricted by
the finite size of the system simulated. However, the fourth-order modulus
$\left< \Upsilon_4 \right>$ described above does not have this problem, as
shown in Ref.~\cite{petter}.
 
\begin{figure}
\includegraphics[width=0.48\textwidth]{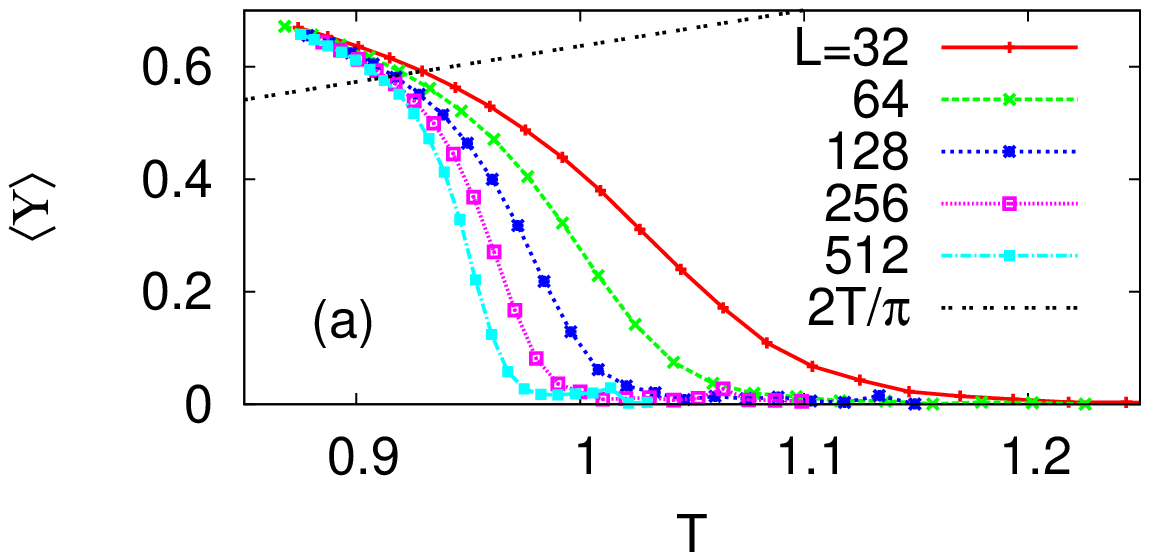}
\includegraphics[width=0.48\textwidth]{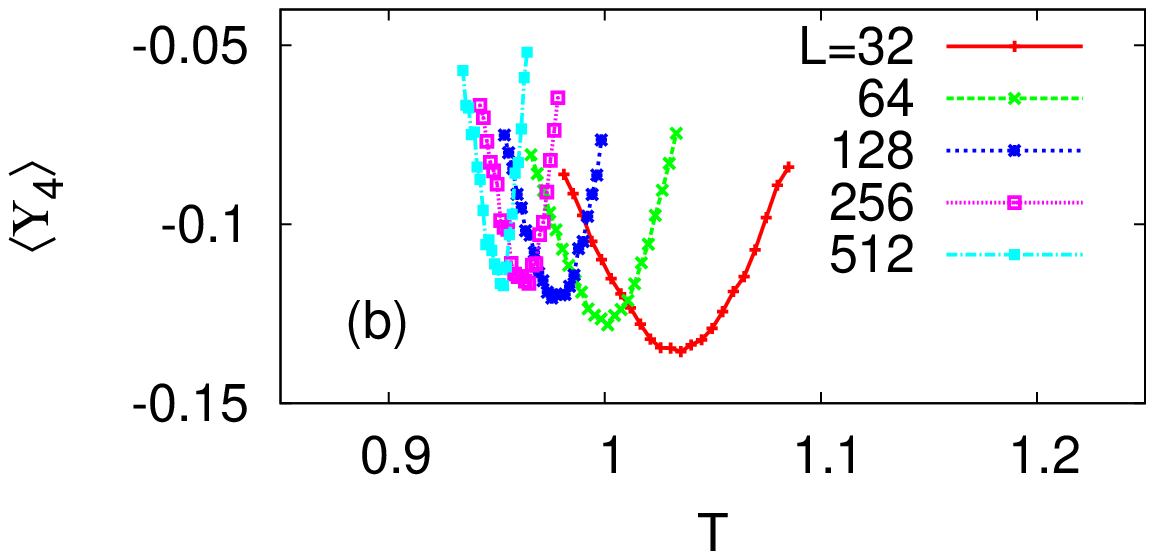}
\includegraphics[width=0.48\textwidth]{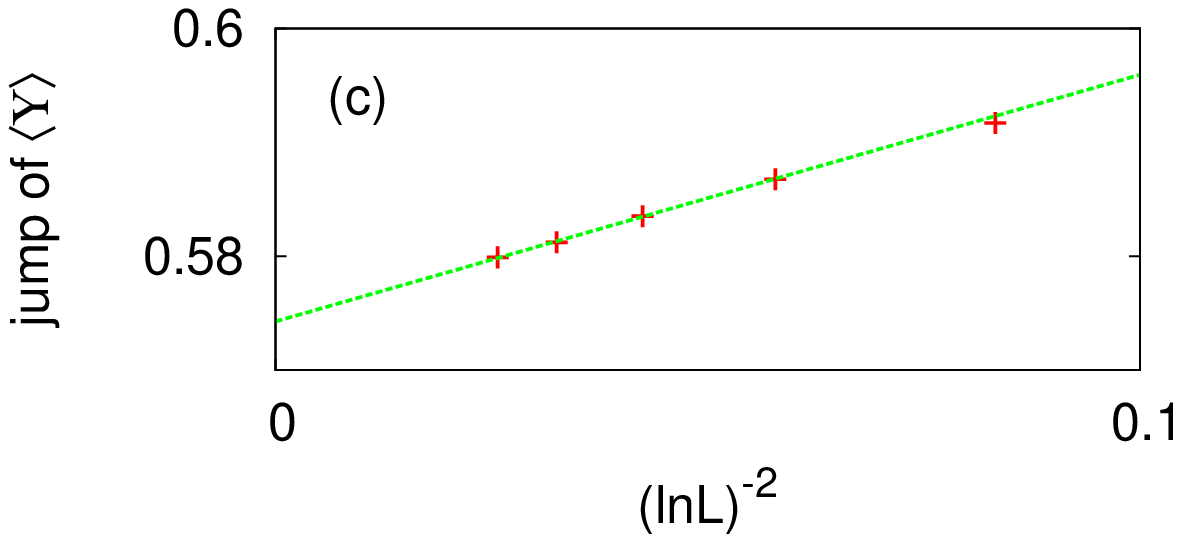}
\includegraphics[width=0.48\textwidth]{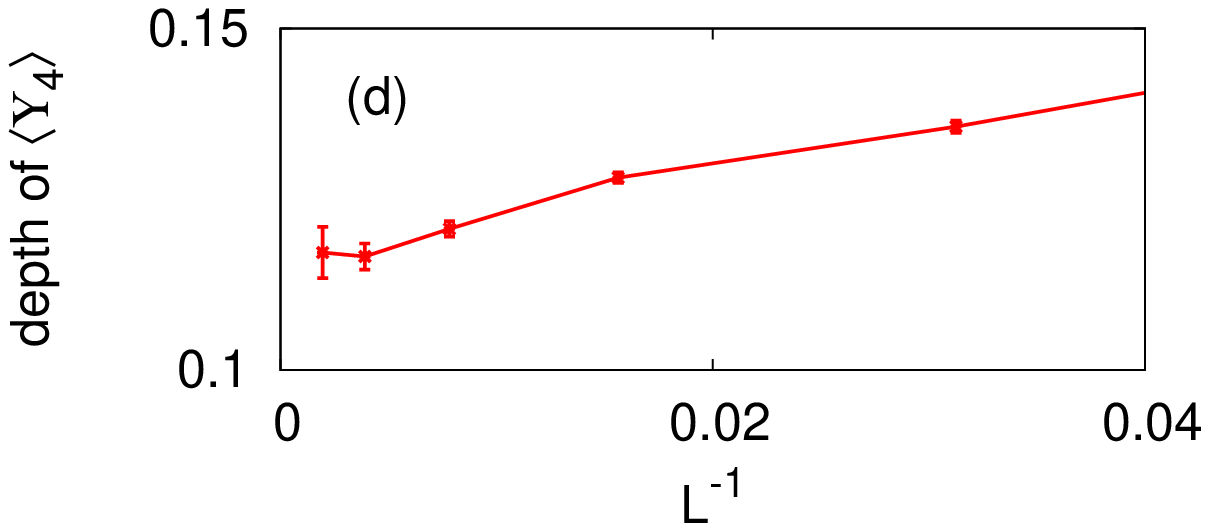}
\caption{(Color online) Temperature dependences of correlation functions,
(a) the helicity modulus and (b) the fourth-order helicity modulus for the
six-state clock model.
(c) Amplitudes satisfying $\left< \Upsilon \right> =
2T/\pi$. The extrapolation yields $\left< \Upsilon \right> =
0.5742(3)$ at $L \rightarrow \infty$.
(d) Size dependence of the depth of $\left< \Upsilon_4 \right>$.
Shown are $L=32, 64, 128, 256$, and $512$ from right to left both in (c) and (d).}
\label{fig:y}
\end{figure}

We have here calculated the helicity modulus and the fourth-order modulus
for the six-state clock model
on $L \times L$ square lattices with the periodic boundary condition
using the Wolff algorithm~\cite{wolff}
[Figs.~\ref{fig:y}(a) and \ref{fig:y}(b)]. One clearly sees
that $\left< \Upsilon_4 \right>$ approaches a negative finite value, which
means that this system will exhibit a discontinuous jump in $\left< \Upsilon
\right>$ at the transition temperature $T_c$ according to the above
argument. In addition,
if it exhibits the universal jump~\cite{jump}, the transition temperature
and the jumping amplitude is related by $\left< Y \right> = 2T_c/\pi$.
Noting that the correlation length $\xi$ scales as $\log\xi \sim \log L \sim
(T-T_c)^{-1/2}$ in the KT
scenario, our extrapolation to $L \rightarrow \infty$ gives
$T_c = 0.9020(5)$ as shown in
Fig.~\ref{fig:y}(c). This agrees well with an independent estimation of
$T_c = 0.9008(6)$ in Ref.~\cite{tomita}, differing only by $0.3\%$ at most.
Recall also that Ref.~\cite{hwang} estimated $T_c$ as $0.997(2)$ ignoring
the KT scenario, which cannot be correct since one finds that the peak of
$\left< \Upsilon_4 \right>$, which signals the criticality, is below this
temperature for large enough $L$ [Fig.~\ref{fig:y}(b)].
These observations rule out the possibility that the upper
transition can be of a non-KT continuous type as suggested in
Ref.~\cite{hwang}.

What could be the reason for the result in Ref.~\cite{hwang}? First of all,
one notes that in Ref.~\cite{hwang} the same critical index $\nu$ is found
for both the upper and lower transitions, suggesting that the two
transitions are identically continuous. However, this is established
only for rather small lattice sizes, $L\leq 28$ used in the study, while
we use lattice sizes up to $L=512$ here. Thus the results found in
Ref.~\cite{hwang} could be an artifact of the small lattice sizes. In
particular, one may notice that for the generalized clock model studied in
Ref.~\cite{baek} the two separate transitions merge into one, when the
cosine interaction is slightly distorted.
At the merging point, the transitions can be described as one joint
continuous transition. Therefore, one possibility could be that the two
transitions for the usual six-state clock model are, for small enough
lattice sizes, both strongly influenced by the critical behavior of the
merging point. If this is the case, then the Fisher-zero approach should
become consistent with a KT transition provided that the lattice sizes are
large enough.

\acknowledgments
S.K.B. and P.M. acknowledge the support from the Swedish Research Council
with the Grant No. 621-2002-4135.
B.J.K. was supported by the Korea Research Foundation Grant funded
by the Korean  Government (MEST) with the Grant No. KRF-2008-005-J00703.
This research was conducted using the resources of High Performance
Computing Center North (HPC2N).


\begin{thebibliography}{1}
\providecommand*{\bibinfo}[2]{#2}
\providecommand*{\eprint}[1]{#1}
\providecommand*{\url}[1]{#1}
\bibitem{hwang}
\bibinfo{author}{C.-O. Hwang}, \bibinfo{journal}{Phys. Rev. E}
  \bibinfo{volume}{\textbf{80}}, \bibinfo{pages}{042103}
  (\bibinfo{date}{2009}).
\bibitem{jkkn}
\bibinfo{author}{J.~V. Jos\'e}, \bibinfo{author}{L.~P. Kadanoff},
  \bibinfo{author}{S.~Kirkpatrick}, and \bibinfo{author}{D.~R. Nelson},
  \bibinfo{journal}{Phys. Rev. B} \bibinfo{volume}{\textbf{16}},
  \bibinfo{pages}{1217} (\bibinfo{date}{1977}).
\bibitem{elit}
\bibinfo{author}{S.~Elitzur}, \bibinfo{author}{R.~B. Pearson}, and
  \bibinfo{author}{J.~Shigemitsu}, \bibinfo{journal}{Phys. Rev. D}
  \bibinfo{volume}{\textbf{19}}, \bibinfo{pages}{3698} (\bibinfo{date}{1979}).
\bibitem{tomita}
\bibinfo{author}{Y.~Tomita} and \bibinfo{author}{Y.~Okabe},
  \bibinfo{journal}{Phys. Rev. B} \bibinfo{volume}{\textbf{65}},
  \bibinfo{pages}{184405} (\bibinfo{date}{2002}).
\bibitem{lapilli}
\bibinfo{author}{C.~M. Lapilli}, \bibinfo{author}{P.~Pfeifer}, and
  \bibinfo{author}{C.~Wexler}, \bibinfo{journal}{Phys. Rev. Lett.}
  \bibinfo{volume}{\textbf{96}}, \bibinfo{pages}{140603}
  (\bibinfo{date}{2006}).
\bibitem{petter}
\bibinfo{author}{P.~Minnhagen} and \bibinfo{author}{B.~J. Kim},
  \bibinfo{journal}{Phys. Rev. B} \bibinfo{volume}{\textbf{67}},
  \bibinfo{pages}{172509} (\bibinfo{date}{2003}).
\bibitem{wolff}
\bibinfo{author}{U.~Wolff}, \bibinfo{journal}{Phys. Rev. Lett.}
  \bibinfo{volume}{\textbf{62}}, \bibinfo{pages}{361} (\bibinfo{date}{1989}).
\bibitem{jump}
\bibinfo{author}{P.~Minnhagen} and \bibinfo{author}{G.~G. Warren},
  \bibinfo{journal}{Phys. Rev. B} \bibinfo{volume}{\textbf{24}},
  \bibinfo{pages}{2526} (\bibinfo{date}{1981}).
\bibitem{baek}
\bibinfo{author}{S.~K. Baek}, \bibinfo{author}{P.~Minnhagen}, and
  \bibinfo{author}{B.~J. Kim}, \bibinfo{journal}{Phys. Rev. E}
  \bibinfo{volume}{\textbf{80}}, \bibinfo{pages}{060101(R)}
  (\bibinfo{date}{2009}).

\end{thebibliography}

\end{document}